\documentclass{PoS}
\usepackage{graphicx}
\usepackage{wasysym}

\title{The effect of $^{12}$C + $^{12}$C rate uncertainties on the weak s-process component}

\ShortTitle{$^{12}$C + $^{12}$C rate uncertainties and the s process}

\author{\speaker{Michael E. Bennett}$^a$, Raphael Hirschi$^{ab}$, Marco Pignatari$^{gcd}$, Steven Diehl$^e$, Chris Fryer$^f$, Falk Herwig$^g$, William Hillary$^g$, Aimee Hungerford$^f$, Debra Richman$^g$, Gabriel Rockefeller$^f$, Frank X. Timmes$^h$, Michael Wiescher$^c$. \\
\llap{$^a$} Astrophysics Group, Keele University, ST5 5BG, UK \\
\llap{$^b$} IPMU, University of Tokyo, Kashiwa, Chiba 277-8582, Japan \\
\llap{$^c$} Joint Institute for Nuclear Astrophysics, University of Notre Dame, IN, 46556, USA \\
\llap{$^d$} TRIUMF, 4004 Wesbrook Mall, Vancouver, BC, Canada, V6T 2A3 \\
\llap{$^e$} Theoretical Astrophysics (T-6), LANL, Los Alamos, NM, 87545, USA \\
\llap{$^f$} Computational Physics and Methods (CCS-2), LANL, Los Alamos, NM, 87545, USA \\
\llap{$^g$} Dept. of Physics \& Astronomy, Victoria, BC, V8W 3P6,Canada \\
\llap{$^h$} School of Earth and Space Exploration, University of Arizona, Tempe, AZ 85287, USA \\
Email: \email{meb@astro.keele.ac.uk}}%, \email{Hirschi@astro.keele.ac.uk}, \email{mpignatari@gmail.com}, \email{clfreyer@lanl.gov}, \email{fherwig@uvic.ca}, \email{aimee@lanl.gov}, \email{gaber@lanl.gov}, \email{fxt44@mac.com}, \email{Michael.C.Wiescher.1@nd.edu}}

\abstract{The contribution by massive stars ($M > 9M_{\odot}$) to the weak s-process component of the solar system abundances is primarily due to the $^{22}$Ne neutron source, which is activated near the end of helium-core burning.  The residual $^{22}$Ne left over from helium-core burning is then reignited during carbon burning, initiating further s-processing that modifies the isotopic distribution.  This modification is sensitive to the stellar structure and the carbon burning reaction rate.  Recent work on the $^{12}$C + $^{12}$C reaction suggests that resonances located within the Gamow peak may exist, causing a strong increase in the astrophysical S-factor and consequently the reaction rate.  To investigate the effect of an increased rate, $25 M_{\odot}$ stellar models with three different carbon burning rates, at solar metallicity, were generated using the Geneva Stellar Evolution Code (GENEC) with nucleosynthesis post-processing calculated using the NuGrid Multi-zone Post-Processing Network code (MPPNP).  The strongest rate caused carbon burning to occur in a large convective core rather than a radiative one.  The presence of this large convective core leads to an overlap with the subsequent convective carbon-shell, significantly altering the initial composition of the carbon-shell.  In addition, an enhanced rate causes carbon-shell burning episodes to ignite earlier in the evolution of the star, igniting the $^{22}$Ne source at lower temperatures and reducing the neutron density.}

\FullConference{11th Symposium on Nuclei in the Cosmos, NIC XI\\
		July 19-23, 2010\\
		Heidelberg, Germany}

\begin{document}

\section{Introduction}

The s-process components identified to contribute to the solar abundance distribution are the weak component, that is produced in massive stars (M > 9$M_{\odot}$), and the main and strong components, that are produced in AGB stars.  In particular, the weak s-process component is responsible for most of the isotopes in the mass range $60 < A < 90$.  During helium-core burning in massive stars, $^{22}$Ne is formed from $^{14}$N synthesized by the CNO cycle, via the reaction chain $^{14}$N$(\alpha,\gamma)^{18}$F$(\beta^+)^{18}$O$(\alpha,\gamma)^{22}$Ne.  At the end of helium burning, when the temperature reaches $0.25$ GK ($1$ GK = $10^9$ K), the $^{22}$Ne$(\alpha,$n$)^{25}$Mg reaction becomes efficient, resulting in an s process characterised by an average neutron density $n_n \sim 10^{6}$ n cm$^{-3}$ and a neutron exposure (for a 25$M_{\odot}$ star) $\tau_n \sim 0.2$ mbarn$^{-1}$.  During the advanced stages, convective carbon-shell burning reignites the remaining $^{22}$Ne with a much higher neutron density but lower neutron exposure ($n_n \simeq 10^{11}$ n cm$^{-3}$ and $\tau_n \simeq 0.06$ mbarn$^{-1}$ \cite{1991ApJ...371..665R}).  The s process also occurs during (radiative) carbon-core burning via the $^{13}$C($\alpha$, n)$^{16}$O neutron source \cite{1998ApJ...502..737C}.  However, in standard $25 M_{\odot}$ stars heavy elements synthesized in the core are further processed and are not ejected during the supernova explosion.  Thus they do not contribute to the final yields.  Changes to the $^{12}$C + $^{12}$C reaction have an effect on the s process in massive stars \cite{2010JPhCS.202a2023B}, but a detailed analysis has so far been limited to the effect of a reduced rate due to fusion hindrance \cite{2007PhRvC..76c5802G}, although the consequences of an increased rate have been considered in superbursts on accreting neutron stars in X-ray binaries \cite{2009ApJ...702..660C}.

\section{The $^{12}$C + $^{12}$C reaction}

The $^{12}$C + $^{12}$C rate used in most stellar models is that of Caughlan \& Fowler (1988) \cite{1988ADNDT..40..283C}.  The recommended average S-factor, $S(E)^*$, at low energies is $3 \times 10^{16}$ MeV barn, which corresponds to an approximate average over resonance structures from $E=2.5$ to $6.5$ MeV.  Unfortunately, information on the resonance structures of $^{12}$C + $^{12}$C near the Gamow peak energy $E_0 = 1.5$ MeV is lacking; due to the very low cross-section at these energies ($\ll 1$ nbarn), experiments at these energies are strongly affected by hydrogen and deuterium contamination of the target \cite{2007PhRvL..98l2501S}\cite{2010JPhCS.202a2025S} (see also contribution by F. Strieder et al. in this volume) and resonance structures at low energies are known to be quasimolecular states, which represents a difficult problem in nuclear physics \cite{1997RPPh...60..819B}.  Nevertheless, the presence of resonances at low energies has been predicted \cite{2006PAN....69.1372P} and a resonance within the Gamow window could dominate the S-factor.

\begin{figure}[htb]
   \includegraphics[trim = 35mm 0 40mm 0, clip, width=\textwidth]{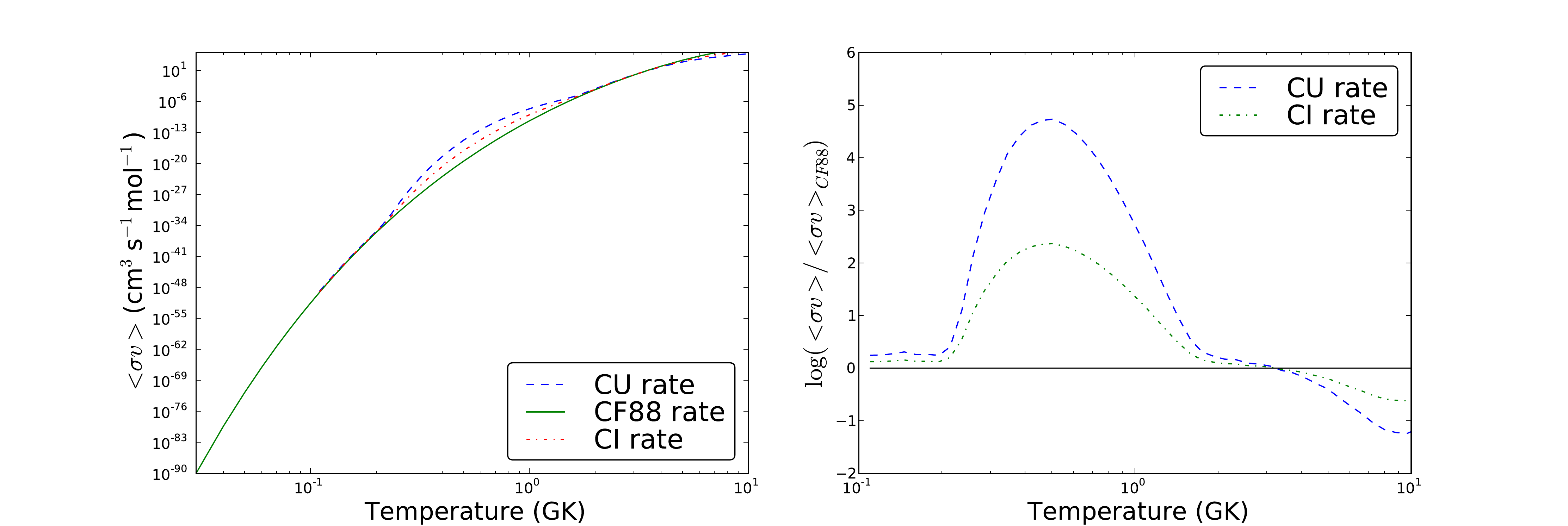}
   \caption{\emph{Left panel}: the cross section of the Caughlan \& Fowler (1988) rate (CF88), upper limit rate (CU) and intermediate rate (CI) as a function of temperature.  \emph{Right panel}: the cross section of the CU and CI rates relative to the CF88 rate.}
   \label{fig:c12rates}
\end{figure}

The three carbon burning rates considered here are the Caughlan \& Fowler (1988) rate (ST), an upper-limit rate (CU) that corresponds to a strong resonance at $E = 1.5$ MeV and an intermediate rate (CI), which is the geometric mean of the standard and upper limit rates (see Fig. \ref{fig:c12rates}).  The enhancement corresponds to a factor $\sim 50,000$ for the upper limit rate and a factor $\sim 250$ for the intermediate rate at a temperature of $0.5$ GK.

\section{Stellar structure and nucleosynthesis}

Non-rotating stellar models of a 25$M_{\odot}$ star were generated for each of three rates in Fig. \ref{fig:c12rates} using the Geneva Stellar Evolution Code (GENEC) \cite{2008ApSS.316...43E}.  The models were post-processed with the parallel variant of the NuGrid Multi-zone Post-Processing Network code (MPPNP) \cite{2008nuco.confE..23H} with a network of $\simeq 1000$ isotopes, up to bismuth.

\begin{figure}[htb]
   \centering
   \includegraphics[trim = 5mm 5mm 5mm 5mm, width=\textwidth]{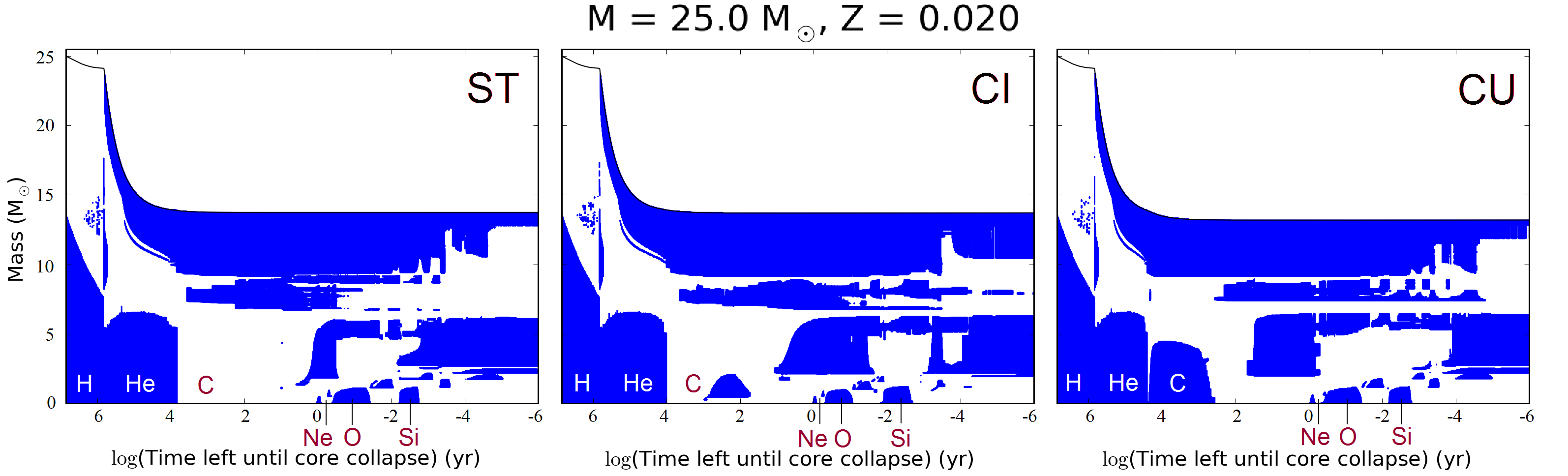}
   \caption{Kippenhahn diagrams for non-rotating stellar models using the CF88 (ST), CI and CU rates (left, centre and right panels respectively).  Shaded regions correspond to convection zones with the main burning stages indicated.}
   \label{fig:kips}
\end{figure}

\begin{figure}[htb]
   \centering
   \includegraphics[width=\textwidth]{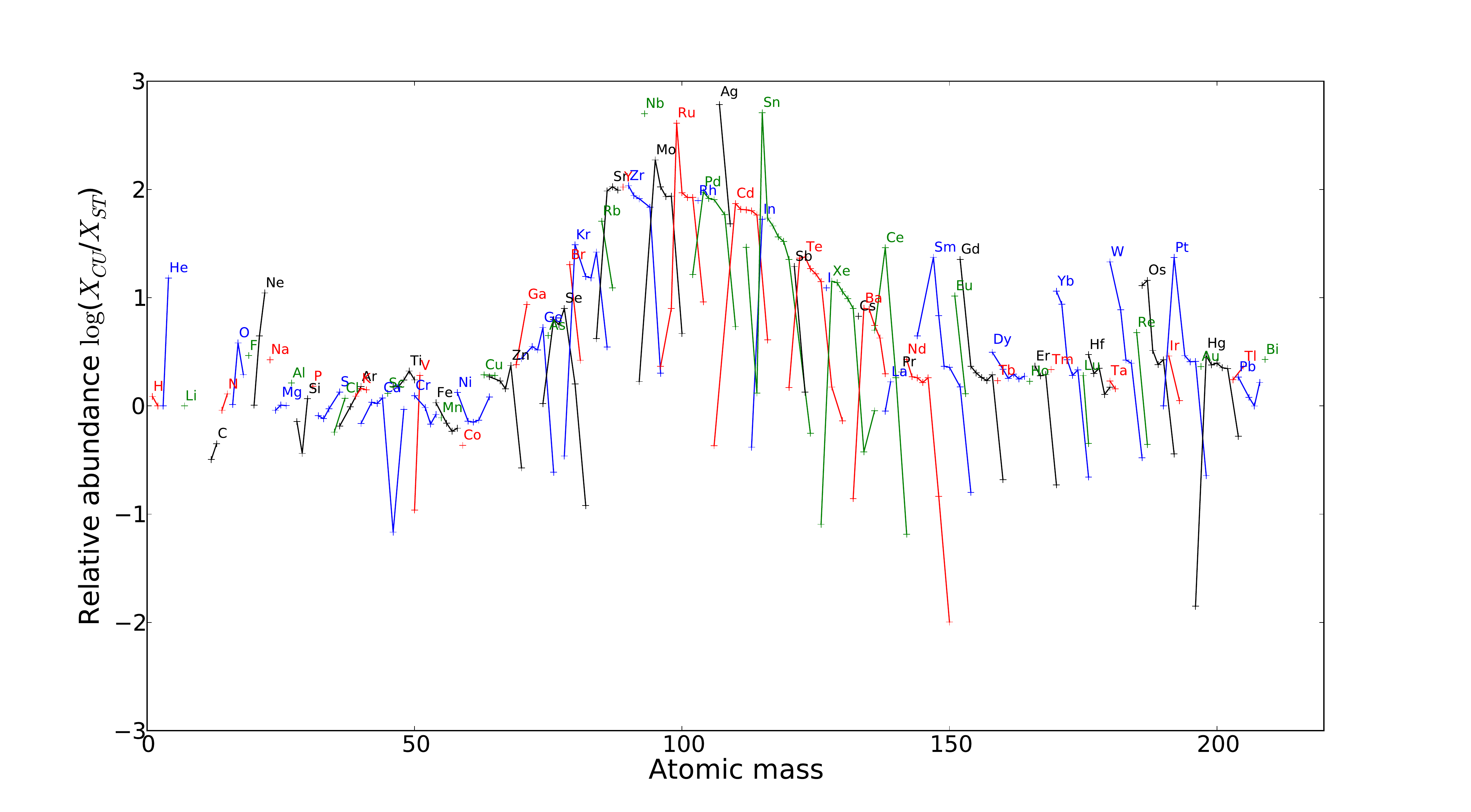}
   \caption{CU model abundances relative to the ST case in the carbon-burning shell.  Lines connect isotopes belonging to the same element.}
   \label{fig:CUoverST}
\end{figure}

\begin{figure}[htb]
   \centering
   \includegraphics[width=\textwidth]{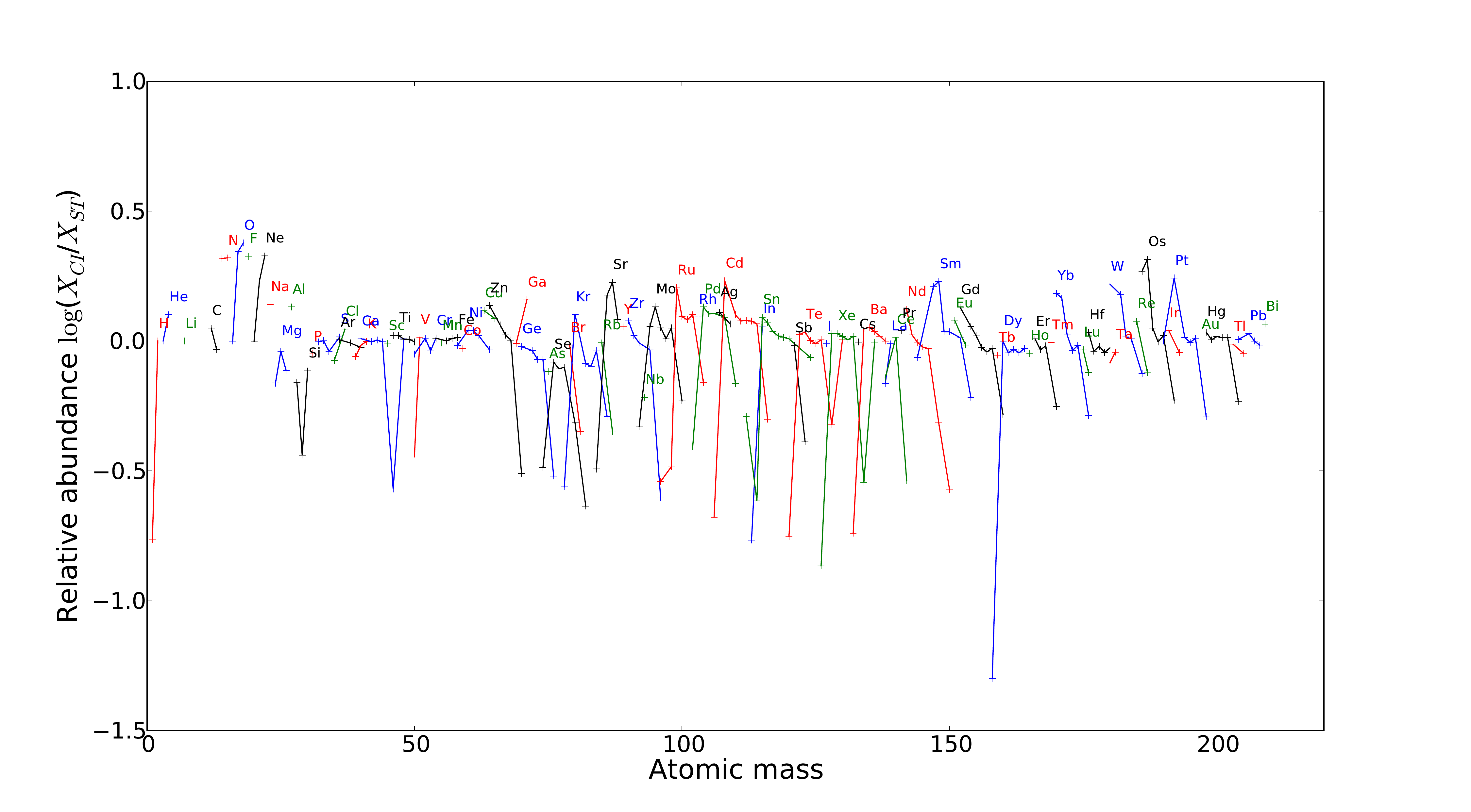}
   \caption{CI model abundances in the second carbon-burning shell relative to ST abundances.  Lines connect isotopes with equal atomic number.  The lower neutron density modifies the isotopic chain, but it should be noted however that some isotopes which appear strongly produced or depleted have very low abundances.}
   \label{fig:CIoverST}
\end{figure}

Figure \ref{fig:kips} shows the Kippenhahn diagrams for the three models.  The CU model features a convective core during carbon-core burning.  Here, the $^{13}$C$(\alpha,$n$)^{16}$O reaction is efficient and provides a neutron exposure comparable to the one in the previous helium-core, increasing significantly the s-process yields due to the previous helium-burning phase.  The convective carbon-core is about $4 M_{\odot}$ and overlaps with the subsequent carbon-shell.  In Fig. \ref{fig:CUoverST} we provide the abundances at the end of the convective carbon-shell compared to the ST case.  The s process powered by the $^{13}$C$(\alpha,$ n)$^{16}$O activation in the carbon-core is strongly efficient and through convective mixing changes the initial composition of the carbon-shell and of the final yields.  Such an effect is particularly evident in the mass range $80 < A < 120$.  Figure \ref{fig:CIoverST} shows the CI model abundances relative to the ST case at the end of the second carbon-shell.  Since the carbon-core is radiative and there is no overlap between the final convective carbon-shell and previous carbon burning events (see Fig. \ref{fig:kips}), the changes in the relative abundances of isotopes are mostly due only to a lower neutron density in the carbon shell, which in turn is caused by the $^{22}$Ne neutron source activating at a lower temperature.  Both the CU and CI models show lowered ignition temperatures (ST: $0.95$GK, CI: $0.74$GK, CU: $0.73$GK) and thus lower neutron densities (ST: $2.02 \times 10^{11}$ n cm$^{-3}$, CI: $2.07 \times 10^{10}$ n cm$^{-3}$,CU: $4.97 \times 10^8$ n cm$^{-3}$).  Notice also that the shells have increased lifetimes in the CU and CI models (ST: $3.4$yr, CI: $10.7$yr, CU: $34.0$yr) and that the neutron exposure in the last convective carbon-shell is similar in each case with a value of $\simeq 0.035$ mbarn$^{-1}$.

To summarise, the presence of a strong resonance in the Gamow window may change the structure and nucleosynthesis of a 25$M_{\odot}$ star, with the main effects being the presence of a convective carbon-core, longer shell-burning lifetimes and decreased ignition temperatures.  Overlap between the convective core and the ensuing shell and lower neutron densities caused by these structural changes will strongly affect the final yields of the star, but firm conclusions should await yields calculations of massive star models at different initial masses, which will be presented in a forthcoming paper (Bennett et al., in prep.).

\acknowledgments{The simulations were generated using the KHAOS cluster at Keele University.}

\end{document}